\documentclass[12pt]{article}
\pdfoutput=1
\usepackage[
colorlinks=true,
linkcolor=blue,
urlcolor=blue,
filecolor=black,
citecolor=red,
]{hyperref}
\usepackage{epsfig}
\usepackage{graphicx, color, xcolor}

\usepackage{amsmath}
\usepackage{amsfonts}
\usepackage{slashed}
\usepackage{amssymb}

\def\ket#1{\mathinner{|{#1}\rangle}}
\def\braket#1{\mathinner{\langle{#1}\rangle}}

\let\protect\relax
{\catcode`\|=\active
  \xdef\Braket{\protect\expandafter\noexpand\csname Braket \endcsname}
  \expandafter\gdef\csname Braket \endcsname#1{\begingroup
     \ifx\SavedDoubleVert\relax
       \let\SavedDoubleVert\|\let\|\BraDoubleVert
     \fi
     \mathcode`\|32768\let|\BraVert
     \left\langle{#1}\right\rangle\endgroup}
}
\def\BraVert{\@ifnextchar|{\|\@gobble}
     {\egroup\,\mid@vertical\,\bgroup}}
\def\BraDoubleVert{\egroup\,\mid@dblvertical\,\bgroup}
\let\SavedDoubleVert\relax

{\catcode`\|=\active
  \xdef\set{\protect\expandafter\noexpand\csname set \endcsname}
  \expandafter\gdef\csname set \endcsname#1{\mathinner
        {\lbrace\,{\mathcode`\|32768\let|\midvert #1}\,\rbrace}}
  \xdef\Set{\protect\expandafter\noexpand\csname Set \endcsname}
  \expandafter\gdef\csname Set \endcsname#1{\left\{%
     \ifx\SavedDoubleVert\relax \let\SavedDoubleVert\|\fi
     \:{\let\|\SetDoubleVert
     \mathcode`\|32768\let|\SetVert
     #1}\:\right\}}
}
\def\midvert{\egroup\mid\bgroup}
\def\SetVert{\@ifnextchar|{\|\@gobble}
    {\egroup\;\mid@vertical\;\bgroup}}
\def\SetDoubleVert{\egroup\;\mid@dblvertical\;\bgroup}

\begingroup
 \edef\@tempa{\meaning\middle}
 \edef\@tempb{\string\middle}
\expandafter \endgroup \ifx\@tempa\@tempb
 \def\mid@vertical{\middle|}
 \def\mid@dblvertical{\middle\SavedDoubleVert}
\else
 \def\mid@vertical{\mskip1mu\vrule\mskip1mu}
 \def\mid@dblvertical{\mskip1mu\vrule\mskip2.5mu\vrule\mskip1mu}
\fi

\makeatletter

\@addtoreset{equation}{section}
\makeatother
\newcommand{\be}{\begin{equation}}
\newcommand{\ee}{\end{equation}}
\newcommand{\ben}{\begin{equation*}}
\newcommand{\een}{\end{equation*}}
\newcommand{\bea}{\begin{eqnarray}}
\newcommand{\eea}{\end{eqnarray}}
\newcommand{\alg}{\begin{align}}
\newcommand{\algx}{\end{align}}

\def\({\left(}
\def\){\right)}
\def\<{\left<}
\def\>{\right>}
\def\!{\right|}
\def\|{\left|}

\def\[{\left[}
\def\]{\right]}

\begin{document}

\begin{titlepage}
\vskip1cm
\begin{flushright}
\end{flushright}
\vskip1.25cm
\centerline{\large
\bf Quantum  Information Metric on $\mathbb{R} \times S^{d-1}$ }
\vskip1cm \centerline{ \textsc{
 Dongsu Bak$^{\, \tt a,c}$,  Andrea Trivella$^{\, \tt b}$} }
\vspace{1cm} 
\centerline{\sl  a) Physics Department,
University of Seoul, Seoul 02504 \rm KOREA}
 \vskip0.3cm
  \centerline{\sl b) Mani L. Bhaumik Institute for Theoretical Physics}
   \centerline{\sl  Department of Physics and Astronomy}
 \centerline{\sl
University of California, Los Angeles, CA 90095
 \rm USA}
  \vskip0.3cm
  \centerline{\sl c)
Center for Theoretical Physics of the Universe}
 \centerline{\sl 
 Institute for Basic Science, Seoul 08826 \rm KOREA}
\vskip0.3cm

 \centerline{
\tt{(dsbak@uos.ac.kr, andrea.trivella@physics.ucla.edu)}} 
  \vspace{1cm}

\centerline{ABSTRACT}
 \vspace{0.75cm} \noindent
We present a formula for the information metric on $\mathbb{R} \times {S}^{d-1}$
for a scalar primary operator of integral dimension $\Delta \, (\,\, > \frac{d+1}{2})$. This formula is checked for various space-time dimensions $d$ and $\Delta$
in the field theory side. We check the formula in the gravity side  using the holographic setup. We clarify the regularization and 
renormalization involved in these computations. We also show that the quantum information metric of an exactly marginal operator agrees with the leading order of the interface free energy of the conformal Janus on Euclidean ${S}^d$, which is checked
for $d=2, 3$. 

\vspace{1.75cm}
\end{titlepage}

\section{Introduction}
It has been proposed \cite{MIyaji:2015mia} that, based on the AdS/CFT correspondence \cite{Maldacena:1997re}, the quantum information metric may serve as a new tool to probe the bulk gravity. This proposed correspondence requires an interesting new dictionary,
which may shed a new light on the issue of bulk decoding from the view point of  field theories. 

The quantum information metric \cite{Caves,Gu}, also known as fidelity susceptibility, measures the distance between two infinitesimally different quantum states. To define this quantity we start by considering a one-parameter family of states $\ket{\Psi (\lambda)}$. The quantum information metric $G_{\lambda \lambda}$ is defined as minus the coefficient of the second order term in the expansion of $|\braket{\Psi (\lambda
	+\delta \lambda) | \Psi (\lambda)} |$ for small $\delta \lambda$:
\bea
|\braket{\Psi (\lambda
	+\delta \lambda) | \Psi (\lambda)} |= 1- G_{\lambda\lambda} \delta \lambda^2 
+ {\cal O}(\delta\lambda^3).
\eea 
The definition can be naturally generalized to the case of a multi dimensional parameter space, even though we will not consider that case in this paper. $\lambda$ could be any sort of parameter that labels a family of quantum states so its nature is very general, in this paper we will specialize to a specific set up in which $\lambda$ is the coupling constant of a local operator $\mathcal{O}$ in the Euclidean signature Lagrangian.

The quantum information metric has applications in understanding quantum phase transitions or  response of a quantum system under some spatially homogeneous perturbations. 
Prime examples  of the 
correspondence \cite{MIyaji:2015mia, Bak:2015jxd, Trivella:2016brw, Alishahiha:2017cuk}  involve a $d$ dimensional CFT perturbed by a scalar primary operator of dimension $\Delta$
and its gravity dual described by the $d+1$ dimensional Euclidean Janus geometry \cite{Bak:2003jk, Bak:2007jm}.  The overlap $|\braket{\Psi (\lambda
	+\delta \lambda) | \Psi (\lambda)} |$ and the corresponding quantum information metric 
can be  realized by a path integral of the CFT on  $\mathbb{R} \times M_{d-1}$ ($M_{d-1}$ being the spatial manifold where the CFT lives) with an operator $\mathcal{O}$ turned on 
whose coupling jumps 
from $\lambda$ to $\lambda +\delta \lambda$ though the interface at $\tau=0$ where $\tau$ is the
Euclidean time coordinate ranged from $[-\infty, \infty]$. The quantum information metric defined in this manner 
diverges in general and needs to be regularized and renormalized, the details of these procedures will be discussed in this paper. 
We note that the quantum information metric will scale extensively as
the spatial volume of the system denoted by ${\rm Vol}{M_{d-1}}$. Below we shall be interested in the quantum information metric only at critical point
where $\lambda=0$. In case of an exactly marginal deformation criticality is preserved at any values of $\lambda$ and one finds that the quantum information metric is $\lambda$ independent. We shall further  
limit our consideration  to the cases of scalar primary operators with the restriction  $\Delta > \frac{d+1}{2}$.

With the choice of $M_{d-1}=\mathbb{R}^{d-1}$ one finds that the (renormalized) quantum information metric
of a scalar primary operator vanishes rather trivially \cite{Trivella:2016brw}. In this note we shall instead consider the quantum information metric on $\mathbb{R} \times {S}^{d-1}$ 
with a scalar primary operator turned on, this set up was first considered in \cite{Trivella:2016brw}.
When $2\Delta-d+1$ is not an even integer, the result  turns out to be finite and independent of RG scale. For 
$2\Delta-d+1$ even the quantum information 
metric involves a logarithmic term which depends on the 
radius of the $d-1$ sphere times the RG scale. Thus the quantum information metric becomes anomalous in this case.  
 We shall present an explicit formula of the quantum information metric for any $d$ and integral $\Delta $ (with $ \Delta >\frac{d+1}{2}$). We shall verify this formula for various cases using both gravity and  field theory computations. 
We find that the degrees of freedom relevant to the information metric can be thought of living in a  $d-1$ dimensional theory
localized on the interface $M_{d-1}$.

In Section \ref{sec:CFT definition} we present the path integral formulation of the quantum information metric including its regularization. 
In Section \ref{sec: field theory computation} we carry out the field theory computation and give the formula for the renormalized quantum information metric.  We check this formula for various cases field theoretically. In Section \ref{sec: holographic checks} we recover the field theory computation from the dual gravity side. This way the holographic regularization scheme  adopted in this note will be justified. In Section \ref{sec: free energy} we relate the quantum information metric to the interface free energy of the conformal Janus  on Euclidean $S^{d}$. 
  Last section 
   is devoted to the concluding remarks and, in Appendix \ref{app},
we explain our normalization of the two point function of operators that is consistent with our gravity description.

\section{Path integral formulation of the quantum information metric}\label{sec:CFT definition}
We firstly review the definition of the quantum information metric on the field theory side. We will use the same formalism and arguments used in \cite{MIyaji:2015mia}.
In particular we focus on the quantum 
information metric for a CFT ground state deformed by a scalar primary operator.
We assume that the CFT lives on a $d$ dimensional cylinder $\mathbb{R}\times {S}^{d-1}$ and that it has an Euclidean signature Lagrangian $\mathcal{L}_0$ of a real scalar field $\Phi$. If one wants to be more general one can think of $\Phi$ as schematically representing all the fundamental fields of the theory.

A generic state $\ket{\, \varphi \,}$ is described by a function  $ \varphi(\Omega)$ on ${S}^{d-1}$, with $\Omega$ being the unit vector in $\mathbb{R}^d$ parameterizing $S^{d-1}$. The overlap between the ground state $\ket{\Psi_0}$ and the generic state $\ket{\,  \varphi \,}$ is obtained by the following Euclidean path integral:
\begin{equation}
\braket{\, \varphi \, |\Psi_0}=\frac{1}{\sqrt{Z_0}} \int_{\Phi(\tau=0, \Omega)=\varphi(\Omega)} \mathcal{D}\Phi \exp\left(-\int_{-\infty}^{0} d\tau \int d^{d-1}\Omega \sqrt{ g_{ {S}^{d-1}}} \mathcal{L}_0\right),
\end{equation}  
where $Z_0$ is the partition function for the theory on the cylinder and  $g_{ {S}^{d-1}}$ is the determinant of the metric of the sphere. For simplicity we set the radius of the sphere $r$ to one, i.e. we measure lengths in units of $r$. We eventually restore factors of $r$ using dimensional analysis.

At this point we can consider deforming the theory by a primary operator $\mathcal{O}$ of conformal dimension $\Delta$. The new Euclidean Lagrangian will be given by 
\begin{equation}
\mathcal{L}_1=\mathcal{L}_0+\delta \lambda \, \mathcal{O}.
\end{equation}
By the same arguments used above the overlap between a generic state $\ket{\,  \varphi \,}$ and the ground state of the new Lagrangian $\mathcal{L}_1$, indicated by $\ket{\Psi_1}$, can be written as:
\begin{equation}
\braket{\,  \varphi \,|\Psi_1}=\frac{1}{\sqrt{Z_1}} \int_{\Phi(\tau=0, \Omega)=\varphi(\Omega)} \mathcal{D}\Phi \exp\left(-\int_{-\infty}^{0} d\tau \int d^{d-1}\Omega \sqrt{ g_{ {S}^{d-1}}} \mathcal{L}_1\right).
\end{equation}  
We can then compute the overlap between the ground state of the original theory and the ground state of the deformed theory. We find
\begin{eqnarray}\label{overlap}
\braket{\Psi_1|\Psi_0}&=&\int \mathcal{D} \varphi \braket{\Psi_1|  \varphi} \braket{ \varphi|\Psi_0}\nonumber\\ 
&=&\frac{\int \mathcal{D} \Phi \exp \left(-\int_{-\infty}^{0} d \tau \int d^{d-1}\Omega \sqrt{ g_{ {S}^{d-1}}}  \mathcal{L}_{0}- \int_{0}^{\infty} d \tau \int d^{d-1}\Omega \sqrt{ g_{ {S}^{d-1}}} \mathcal{L}_1\right)}{(Z_0 Z_{1})^{1/2}}.
\end{eqnarray}
One should regard this result as a formal equation. In fact, because of the sudden change in the action at $\tau=0$, this path integral suffers UV divergences which require regularization and renormalization. We introduce a regulator $\epsilon$ by deforming the state $\ket{\Psi_1}$ in the following way:
\begin{equation}
\ket{\Psi^\epsilon_1 }=\frac{e^{-\epsilon H_0}\ket{\Psi_1}}{\left(\braket{\Psi_1|{e^{-2\epsilon H_0}|\Psi_1}} \right)^{1/2}}.
\end{equation}
This choice makes the path integral formulation of $\braket{\Psi^\epsilon_1|\Psi_0}$ well defined\footnote{There are other ways to regularize this path integral, for example one could also consider to deform the state $\ket{\Psi_0}$. The regularization adopted here is convenient because it induces a nice geometrical regularization for the quantum information metric.}.
One can now perform an expansion of $|\braket{\Psi^\epsilon_1|\Psi_0}|$ for small $\delta \lambda$. The regularized quantum 
information metric is defined as minus the coefficient of the $\delta \lambda^2$ term
\begin{equation}
|\braket{\Psi^\epsilon_1|\Psi_0}|=1-G^{\epsilon}_{\lambda \lambda} 
\, \delta \lambda^2+\mathcal{O}(\delta \lambda^3).
\end{equation}
Using the path integral formulation and applying a perturbative expansion in $\delta \lambda$ one finds
\begin{equation}\label{eq:Gint}
G^{\epsilon}_{\lambda \lambda} =\frac{1}{2}\int d^{d-1}\Omega_1 \sqrt{g_{ {S}^{d-1}}} \int d^{d-1}\Omega_2 \sqrt{g_{ {S}^{d-1}}}\int_{-\infty}^{-\epsilon} d \tau_1 \int_{\epsilon}^{\infty}d \tau_2\braket{\mathcal{O}(\tau_1,\Omega_1)\mathcal{O}(\tau_2,\Omega_2)}.
\end{equation}
The regularization procedure effectively removes a strip shaped region centered on $\tau=0$.
\section{Field theory computation}\label{sec: field theory computation}
To compute $G^\epsilon_{\lambda \lambda} $ we need to use the two point function for a primary operator on the cylinder. We start with the two point function for $\mathbb{R}^{d}$ in Euclidean signature:
\begin{equation}
\braket{\mathcal{ O}(\tau_P, x) \mathcal{O}(\tau_P', x')}  = \frac{{\cal N}_{\Delta}} {\left[ \, (\tau_P-\tau'_P)^2 +(x-x')^2 \,  \right]^\Delta}
\end{equation} 
where $\tau_P$ indicated the Euclidean time. We choose the following normalization constant
\begin{equation}\label{normalization}
{\cal N}_\Delta= \frac{2 \eta\ \ell^{d-1}\, d \, \Gamma(\Delta ) }{\pi^{\frac{d }{2}} \Gamma(\Delta -\frac{d}{2})  }
\end{equation}
where $\eta= \frac{1}{16\pi G}$, with the $d+1$ dimensional Newton's constant $G$, and $\ell$ is the AdS radius scale appearing in the dual gravity description. 
This normalization is used to guarantee agreement between bulk and field theory side. We give a more detailed discussion and a derivation of this relation in Appendix \ref{app}.

\begin{figure}[t!]
	\centering  
	\includegraphics[width=8cm]{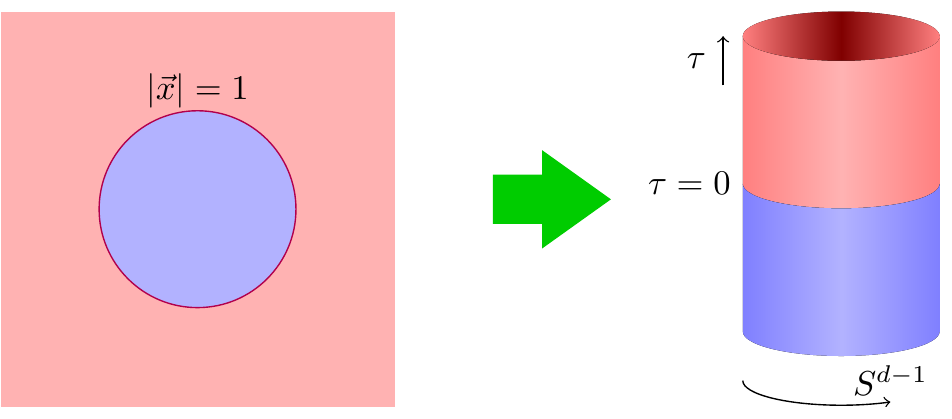}
	\caption{\small  A map from $\mathbb{R}^d$ to $\mathbb{R} \times S^{d-1}$.
	}
	\label{fig1}
\end{figure}

Since the metric of Euclidean signature $\mathbb{R}^d$ 
\begin{equation}
ds^2=d\tau_P^2+\sum_{i}^{d-1}(dx^i)^2=d\xi^2+\xi^2 ds^2_{ {S}^{d-1}}
\end{equation}
and the metric of the cylinder
\begin{equation}
ds^2=d\tau^2+ ds^2_{{S}^{d-1}}
\end{equation}
are related by the conformal transformation $\xi=\exp(\tau)$, we can easily find the following expression for the two point function on the cylinder
\begin{equation}
\braket{\mathcal{O}(\tau_1,\Omega_1)\mathcal{O}(\tau_1,\Omega_2}=\frac{{\cal N}_\Delta}{(2 \cosh(\tau_1-\tau_2)-2 \Omega_1 \cdot \Omega_2)^{\Delta}}.
\end{equation}
We depict the corresponding conformal map in Fig.~\ref{fig1}.
The form of the two point function implies that in the $\epsilon \rightarrow 0$ limit one gets the following leading behavior for the quantum 
information metric
\begin{equation}
G^\epsilon_{\lambda \lambda} \approx \epsilon^{d-2\Delta+1}.
\end{equation}
This is not a surprise. In fact we recover the same leading behavior as the case of a CFT living in flat space \cite{MIyaji:2015mia,Bak:2015jxd,Trivella:2016brw}.

What makes the configuration of the cylinder more interesting is the existence of a physical universal contribution. In addition, even if flat space and the cylinder are conformally equivalent, the quantum 
information metric on the cylinder cannot be inferred in general  by the knowledge of the quantum 
information metric in flat space. This is due to the fact that we are turning on dimension-full coupling constants in the path integral formulation which results in the breaking of conformal symmetry.

In the following we focus on integer values of the conformal dimension $\Delta$ and we take $\Delta > (d+1)/2$ to avoid the issue of infrared divergences. We now start to work on the integral appearing in equation (\ref{eq:Gint}).  We fix $\Omega_2$ and we integrate over $\Omega_1$. Since $\Omega_2$ is fixed we can take it as the north pole for the coordinates system used in the $\Omega_1$ integration. We then have
 \begin{equation}
 \int d^{d-1}\Omega_1 \sqrt{g_{S^{d-1}}}\braket{\mathcal{O}(\Omega_1,\tau_1)\mathcal{O}(\Omega_2,\tau_2)}={\cal N}_\Delta \int_{0}^{\pi} d\theta \frac{\sin{\theta}^{d-2}\,\mathrm{Vol}{S^{d-2}}}{(2 \cosh(\tau_1 -\tau_2)-2 \cos{\theta})^{\Delta}}.
 \end{equation}
 The integral 
 \begin{equation}
 \mathcal{I}=\int_{0}^{\pi} d\theta \frac{ \sin{\theta}^{d-2}}{(2 \cosh(\tau_1-\tau_2)-2 \cos{\theta})^{\Delta}}
 \end{equation}
 can be performed and it produces the following result:
 \begin{equation}\label{eq:integrand}
 \mathcal{I}=\begin{cases}
 \frac{\pi ^{3/2} 2^{-\Delta } (-1)^{n+1} (\cosh (\tau_1-\tau_2)+1)^{-\Delta } \, _2F_1\left(n-\frac{1}{2},\Delta ;2 n-1;\frac{2}{\cosh (\tau_1-\tau_2)+1}\right)}{\Gamma \left(\frac{3}{2}-n\right) \Gamma (n)} & d=2n\\
 \frac{\sqrt{\pi } 2^{-\Delta } (n-1)! (\cosh (\tau_1-\tau_2)+1)^{-\Delta } \, _2F_1\left(n,\Delta ;2 n;\frac{2}{\cosh (\tau_1-\tau_2)+1}\right)}{\Gamma \left(n+\frac{1}{2}\right)}& d=2n+1
 \end{cases}
 \end{equation}
 We can use the fact that $\mathcal{I}$ depends only on the difference $\tau_1-\tau_2$ to simplify the form of $G_{\lambda \lambda}^{\epsilon}$:
 \begin{equation}
 G^\epsilon_{\lambda \lambda}
 =\frac{1}{2}{\cal N}_\Delta\mathrm{Vol}{S^{d-2}}\mathrm{Vol}{S^{d-1}} \int_{\epsilon}^{\infty} d\tau_1 \int_{-\infty}^{-\epsilon} d\tau_2 \mathcal{I}(\tau_1-\tau_2).
 \end{equation}
 At this point we change variables. We introduce $u=\tau_1-\tau_2$ and $v=\tau_1+\tau_2$. The Jacobian give a factor of $1/2$. We are then left with
 \begin{eqnarray}\label{eq:G-as-integral}
 G_{\lambda \lambda}^\epsilon&=&\frac{1}{4}{\cal N}_\Delta\mathrm{Vol}{S^{d-2}}\mathrm{Vol}{S^{d-1}}  \int_{2 \epsilon}^{\infty} du \int_{-u+2 \epsilon}^{u-2 \epsilon} dv  \, \mathcal{I}(u)\nonumber\\
 &=&\frac{1}{2}{\cal N}_\Delta\mathrm{Vol}{S^{d-2}}\mathrm{Vol}{S^{d-1}}  \int_{2 \epsilon}^{\infty} du ( u -2 \epsilon)\,   \mathcal{I}(u).
 \end{eqnarray}
As $\epsilon \rightarrow 0$ $G_{\lambda \lambda}^\epsilon$ admits the following expansion:
\bea
G^\epsilon_{\lambda \lambda}= a_{-2\Delta +d+1} \left(\frac{\epsilon}{r}\right)^{-2\Delta +d+1}+  a_{-2\Delta +d-1} \left(\frac{\epsilon}{r}\right)^{-2\Delta +d-1} + \cdots
+ a_0 + b_0 \log \frac{\epsilon}{r}+ {\cal O}(\epsilon)
\eea
where we have restored the radius $r$ of the spatial sphere where the CFT lives. The logarithmic term is present only when $2 \Delta -d-1$ is even.

To extract the universal piece one has in general to construct counterterms that need to be added to action. This is a standard procedure in QFT. We choose to work in the minimal subtraction scheme. Once the power divergences are removed we can identify the universal piece in

 \bea
 G_{\lambda \lambda}= \left\{ 
 \begin{array}{cc}
 	 -b_0 \log \mu r &    \ {\rm if} \ 2 \Delta -d-1 \ {\rm is} \ {\rm even}\\
 	a_0 &   {\rm otherwise.}~~~~~~~~~~~~~~
 \end{array}
 \right.
 \eea
 where $\mu$ is the renormalization scale.
 This is can be explained heuristically when $\Delta=d$. In fact the path integral formulation can be interpreted as the partition function of a field theory with a conformal defect. The conformal defect lives in $d-1$ dimension, it is not a surprise that the anomalous term (logarithmic divergence) appears for $d$ odd.
 
The computation of $G_{\lambda \lambda}$ is in principle a well posed problem and it is easy to work on specific cases, however it seems that a generic derivation of $G_{\lambda \lambda}$ is difficult to obtain. Based on numerous checks we propose that the universal contribution of the quantum 
information metric for a CFT living on the cylinder deformed by a scalar primary operator is given by
\begin{itemize}
\item $d$ even: 
\bea\label{eq:G-d-even}
G_{\lambda\lambda}=\eta \ell^{d-1} \frac{d}{4} (-1)^{[\Delta -\frac{d-1}{2}]} \frac{\[\Gamma(\frac{\Delta}{2})\Gamma(\frac{\Delta}{2}-\frac{d-2}{2})\]^2}{\Gamma(\Delta-\frac{d}{2})\Gamma(\Delta-\frac{d-2}{2})} \mathrm{Vol}{S^{d-1}}
\eea 
where $\mathrm{Vol}{S^{d-1}}$ is the volume of unit $S^{d-1}$ given by
\bea
\mathrm{Vol}{S^{d-1}}= \frac{2 \pi^{\frac{d}{2}}}{\Gamma(\frac{d}{2})}
\eea
\item $d$ odd:
\bea\label{eq:G-d-odd}
G_{\lambda\lambda}&=&G_{\lambda\lambda}^{\mathrm{log}} \log \mu r \nonumber \\
G_{\lambda\lambda}^{\mathrm{log}}&=&=\eta \ell^{d-1} \frac{d}{4} (-1)^{[\Delta -\frac{d-1}{2}]} 
\frac{\[\Gamma(\frac{\Delta}{2})\Gamma(\frac{\Delta}{2}-\frac{d-2}{2})\]^2}
{\Gamma(\Delta-\frac{d}{2})\Gamma(\Delta-\frac{d-2}{2})} \mathrm{Vol}{S^{d-1}} \left(-\frac{2}{\pi}\right).
\eea 

\end{itemize}
\subsection{Checks}\label{checks}
We now perform some explicit checks to validate the suggested formulas. We start by discussing in details a couple of specific examples to show how the computations can be carried out. We then present a list of cases used to check the claims of equations (\ref{eq:G-d-even}) (\ref{eq:G-d-odd}). 
\begin{itemize}
\item $d=2$: Let us start by considering the explicit example $\Delta=3$. If we plug these values in equations (\ref{eq:integrand}) and (\ref{eq:G-as-integral}) we get the following expression
\begin{equation}
G^\epsilon_{\lambda \lambda}={\mathcal{N}_{3}}\pi \int_{2 \epsilon }^{\infty} du \frac{1}{16} \pi  (u-2 \epsilon ) (\cosh 2 u +2) \, \text{csch}^5u.
\end{equation}
This integral can be performed analytically. We can then expand $G_{\lambda \lambda}^{\epsilon}$ in a Laurent series in $\epsilon$, we find
\begin{equation}
G^\epsilon_{\lambda \lambda}=2{\mathcal{N}_{3}}\pi\left(\frac{\pi }{512 \epsilon ^3}-\frac{\pi }{128 \epsilon }+\frac{\pi ^3}{512}+\mathcal{O}(\epsilon) \right).
\end{equation}
The universal contribution is then given by
\begin{equation}
G_{\lambda \lambda}=\frac{\eta \ell \pi^3}{32},
\end{equation}
where we used $\mathcal{N}_{3}=8\eta \ell/\pi$. This matches equation (\ref{eq:G-d-even}).

The same strategy can be applied to other values of $\Delta$. Here is a list of results obtained:
\begin{eqnarray}\label{eq:QIM d=2 examples}
G_{\lambda \lambda}(\Delta=4)&=&-\frac{\pi  \eta  \ell}{12}  \nonumber \\
G_{\lambda \lambda}(\Delta=5)&=& \frac{\pi^3  \eta  \ell}{256}  \nonumber \\
G_{\lambda \lambda}(\Delta=6)&=&-\frac{\pi  \eta  \ell}{180}  \nonumber \\
G_{\lambda \lambda}(\Delta=7)&=&\frac{75 \pi ^3 \eta  \ell}{524288}.
\end{eqnarray}
Equation (\ref{eq:G-d-even}) correctly reproduces all these results.

\item $d=3$ and generic $\Delta$: Let us focus on $d=3$ on $\Delta$ integer
\begin{eqnarray}
G_{\lambda \lambda}&=&\frac{1}{2}\mathcal{N}_{\Delta}\mathrm{Vol}{S_{d-2}}\mathrm{Vol}{S_{d-1}} J \\
J&=&\int_{2 \epsilon}^{\infty}du (u-2\epsilon)\mathcal{I}(u) \\
\mathcal{I}(u)&=&\frac{2^{-\Delta } (\cosh u-1)^{-\Delta } \left(-(\cosh u +1) \tanh ^{2 \Delta } \frac{u}{2}+\cosh u-1\right)}{\Delta -1}.
\end{eqnarray}
We change variable introducing $z=\tanh^2 \frac{u}{2}$. This produces
\begin{eqnarray}
J&=&\epsilon J_1+J_2 \nonumber\\
J_1&=&\int_{\tanh^2\epsilon}^{1}\frac{2^{2-2 \Delta } \left(\frac{1}{z}-1\right)^{\Delta } \left(z^{\Delta }-z\right) }{(\Delta -1) (z-1)^2 \sqrt{z}}dz \nonumber\\
J_2&=&\int_{\tanh^2\epsilon}^{1}\frac{2^{1-2 \Delta } \left(\frac{1}{z}-1\right)^{\Delta } \left(z-z^{\Delta }\right) 
\cosh ^{-1}
\frac{z+1}{1-z}
}{(\Delta -1) (z-1)^2 \sqrt{z}}dz.
\end{eqnarray}
We can proceed as before. If $\Delta$ is integer we have that $J_2$ has a logarithmic divergence while $\epsilon J_1$ does not. So we focus on $J_2$. The logarithmic divergence of $J_2$ corresponds to the coefficient of the $\epsilon^{-1}$ divergence in $\partial_\epsilon J_2$. We have
\begin{equation}
\partial_\epsilon J_2=\frac{2^{3-2 \Delta } \epsilon  \left(\tanh ^{2 \Delta -2} \epsilon -1\right) \text{csch}^{2 \Delta -2} \epsilon }{\Delta -1}.
\end{equation}
The only term that has a $\epsilon^{-1}$ divergence is
\begin{equation}
-\frac{2^{3-2 \Delta } \epsilon  \text{csch}^{2 \Delta -2} \epsilon }{\Delta -1}.
\end{equation}
So we have the coefficient of the log divergence as
\begin{equation}
G_{\lambda \lambda}^{\text{log}}= 4 \pi^3\mathcal{N}_{\Delta} \text{Res}\left(-\frac{2^{3-2 \Delta } \epsilon  \text{csch}^{2 \Delta -2} \epsilon  }{\Delta -1}\right)\bigg|_{\epsilon=0} .
\end{equation}
Here is a list of cases:
\begin{eqnarray}
G_{\lambda \lambda}^{\mathrm{log}}(\Delta=3)&=& 4 \pi \eta \ell^2 \nonumber \\
G_{\lambda \lambda}^{\mathrm{log}}(\Delta=4)&=&-\frac{16 \pi \eta \ell^2}{15 } \nonumber \\
G_{\lambda \lambda}^{\mathrm{log}}(\Delta=5)&=& \frac{16 \pi \eta \ell^2}{175 }  \nonumber \\
G_{\lambda \lambda}^{\mathrm{log}}(\Delta=6)&=&-\frac{256 \pi \eta \ell^2}{3675 } \nonumber \\
G_{\lambda \lambda}^{\mathrm{log}}(\Delta=7)&=&\frac{256 \pi \eta \ell^2}{14553}.
\end{eqnarray}
Notice that equation (\ref{eq:G-d-odd}) correctly reproduces all these results.

\end{itemize}

\subsection{Marginal deformation}
If the primary operator used to deform the theory is an exactly marginal operator ($\Delta=d$) equations (\ref{eq:G-d-even}) and (\ref{eq:G-d-odd}) reduce to
\begin{equation}\label{eq:QIM marginal}
G_{\lambda\lambda}=\begin{cases}
\frac{1}{2}\eta \ell^{d-1} (-1)^{\frac{d}{2}} \mathrm{Vol}{S^{d-1}} & d\phantom{aa} \text{even} \\ 
\frac{1}{\pi}\eta \ell^{d-1} (-1)^{\frac{d-1}{2}} \mathrm{Vol}{S^{d-1}} {\log \mu r }  & d\phantom{aa} \text{odd}.
\end{cases}
\end{equation}
We have checked these results explicitly for $d=2,...,8$ using the same approach adopted in Section \ref{checks}.

\section{Holographic checks}\label{sec: holographic checks}
In this section we firstly review the holographic set up for the computations of the quantum
information metric. We then proceed to examine some explicit examples.
\subsection{Holographic Formulation}
We can write equation (\ref{overlap}) as
\begin{equation}
\braket{\Psi_1|\Psi_0}=\frac{Z_2}{\sqrt{Z_1 Z_0}}
\end{equation}
where $Z_0$ is the partition function of the undeformed CFT, $Z_1$ is the partition function of the theory obtained by deforming the original CFT with a primary scalar operator, $Z_2$ is the partition function of a theory obtained deforming the original CFT only for $\tau>0$.

These quantities can be computed holographically by computing the on shell action of an Einstein-scalar theory with negative cosmological constant.
We work in the large $N$ approximation, where the bulk theory is classical. In principle we would have to solve the equations of motion asking that the metric is asymptotically AdS, i.e. for large $u$ the metric approaches
\begin{equation}
ds^2=\left(1+\frac{u^2}{\ell^2}\right)d\tau^2+\frac{du^2}{1+\frac{u^2}{\ell^2}}+u^2 d\Omega_{d-1}^2+...
\end{equation}
(where the subleading terms start at order $\mathcal{O}(u^{-1})$) and that the scalar field dual to an operator ${\cal O}$ in the field theory side obeys  the following boundary condition
\begin{equation}
\lim_{u\rightarrow \infty} u^{d-\Delta} \phi(u,\tau,\Omega)=\delta \lambda s_k(\tau)
\end{equation}
where 
\begin{eqnarray}
s_0(\tau)&=&0\nonumber \\
s_1(\tau)&=&1 \nonumber \\
s_2(\tau)&=&\begin{cases}
1&\text{if}\phantom{aa}\tau\ge 0\\
0&\text{if}\phantom{aa}\tau\leq 0.
\end{cases}
\end{eqnarray}
The subscript indicates what boundary condition $s_k(\tau)$ needs to be chosen for the construction of $Z_k$, with $k=0,1,2$.

Since in our computation we are only interested in infinitesimal $\delta \lambda$ we can perform a perturbative analysis whose detailed explanation
 can be found in \cite{Trivella:2016brw}. 
One finds
\begin{equation}
-\log Z_k=I_{AdS}+\delta I_k+{\cal O}(\delta \lambda^4),
\end{equation}
where $I_{AdS}$ is the on shell action of pure Einstein theory with negative cosmological constant and $\delta I_k$ is the on shell action for scalar fields probing a fixed AdS background. In particular
\begin{eqnarray}
\delta I_k&=&\eta\int_{\partial {\mathcal{M}}_\epsilon}\sqrt{\gamma_0} n_{\mu}\,  g^{\mu \nu} \phi_k \partial_\nu \phi_k,
\end{eqnarray}
where ${\mathcal{M}}_\epsilon$ is the regularized version of AdS, $\gamma_0$ is the determinant of the induced metric at the 
cut-off surface $\partial \mathcal{M}_\epsilon$ and $n^\mu$ is the unit normal vector at $\partial \mathcal{M}_\epsilon$.   The details of the regularization procedure will be spelled out later.
As was carefully shown  in \cite{Trivella:2016brw}, the matter contributions $\delta I_k$ are  solely responsible for the quantum information metric  while the contributions from the 
metric perturbation are of order  $\delta \lambda^4$ and can be safely ignored.  

The first step in our computation is to find the profile of the scalar field that obeys the equation of motion derived by the following action:
\bea
S=\eta \int_{M_\epsilon}  \sqrt{g_{d+1}} \, (g^{\mu\nu}\nabla_\mu \phi \nabla_\nu \phi + m^2 \phi^2 ).
\eea
To do this we start with Poincar\'e AdS with metric
\bea
ds^2= \frac{\ell^2}{z^2} (dz^2 + dx^i dx^i)
\eea 
with $i=1, \cdots, d$. On this space we can construct a scalar field obeying the equation of motion using the bulk to boundary propagator
\bea
\phi( z, \vec{x}) = c_\Delta \int d^d x' \[ 
\frac{z}{z^2 +(\vec{x}-{\vec{x}}')^2}
\]^\Delta \tilde{s}({\vec{x}}'),
\eea
where
\begin{equation}
c_{\Delta}=\left(\pi^{d/2}\frac{ \Gamma \left(\Delta -\frac{d}{2}\right)}{ \Gamma (\Delta )}\right)^{-1}
\label{cdd}
\end{equation}
and $\tilde{s}({\vec{x}})$ dictates the boundary behavior of the field.
We can reformulate the problem in another system of coordinates where we write $AdS_{d+1}$ in $AdS_{d}$ slicing. The change of coordinates is given by:
\bea
&& z= \frac{\sinh p}{\cosh (y-p)} \cr
&& x^i =  \frac{\cosh y}{\cosh (y-p)}   \Omega^i
\eea
and leads to 
\bea\label{eq:AdS slicing}
ds^2= \ell^2 \left(dy^2 +  \frac{\cosh^2 y}{\sinh^2 p} \left( dp^2 + ds^2_{S^{d-1}} \right)\right)
\eea
where $y \in (-\infty,\infty)$ and $p \in (0, \infty)$. This space is global $AdS_{d+1}$ and we identify its boundary as the Euclidean cylinder $\mathbb{R}\times  {S}^{d-1}$. In general the function $\tilde{s}(x)$ is not invariant under this change of coordinates. From the CFT point of view this is clear: we identify the function $\tilde{s}(x)$ as the coupling constant for the operator $\mathcal{O}$ dual to $\phi$. Since $\mathcal{O}$ is in general not marginal the change of coordinates produces a different coupling constant $s(x)=\tilde s (x) |x|^{d-\Delta}$. This means that if we want to impose a certain boundary condition $s(x)$ on the cylinder we need to choose the boundary condition in flat space to be $\tilde s(x)=s(x)|x|^{\Delta-d}$.

\begin{figure}[t!]
	\centering  
	\includegraphics[width=7cm]{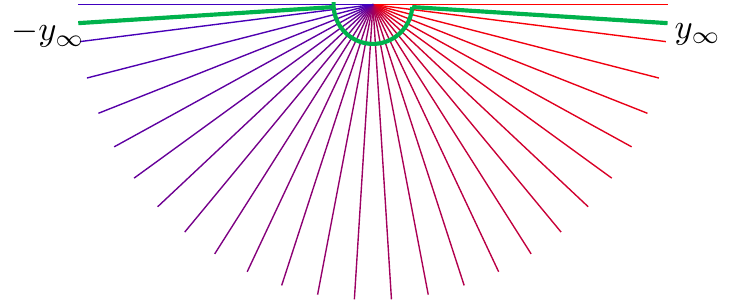}
	\caption{\small   We depict here the regularization  in $(y,p)$ plane of $M_\epsilon$. The regularization is defined by the coordinate ranges
		$y \in (-y_\infty, y_\infty )$ and $p \in (\epsilon, \infty)$. 
	}
	\label{fig3}
\end{figure}

Thus the field $\phi$ used to construct the various partition functions is obtained by performing the following integral
 \bea\label{eq:integration to get the field}
 \phi( z, \vec{x}) = c_\Delta \int d^d x' \[ 
 \frac{z}{z^2 +(\vec{x}-{\vec{x}}')^2}
 \]^\Delta s ({\vec{x}}') |{\vec{x}}'|^{\Delta -d},
 \eea
with $s(x)$ chosen to be 
\bea
&& s_0(\vec{x}) =0 \cr
&& s_1(\vec{x})  =1 \cr
&& s_2(\vec{x}) = \left\{ 
\begin{array}{cc}
	1 &  ~~~~{\rm if} \  |\vec{x}| \ge 1 \\
	0 & ~~~~{\rm otherwise}.
\end{array}
\right.  
\eea 
The map between flat space and the cylinder sends the ball $|\vec{x}|<1$ to the half cylinder $\tau<0$. This explains the form of $\tilde s_2(\vec{x})$. Notice that $\phi_0=0$. Once the field $\phi_k$ has been constructed one needs to proceed to the computation of the on shell action. This quantity is not finite and needs to be regulated. We have seen that the CFT regulator effectively removes the region close to the interface from the path integral. Since we identify the $AdS_{d}$ slices as naturally dual to the interface an obvious bulk regularization is given by taking $p\in (\epsilon, \infty)$. This does not tame all divergences as the integration along the non compact coordinate $y$ will still produce infinities. We then bound $y$ to take value into $(-y_{\infty},y_{\infty})$. The presence of a second cut-off might seem bizarre, however we will notice that the final result will be finite in the $y_{\infty}\rightarrow \infty$ limit. For a more detailed discussion of the two cut-off procedure in holographic computations we refer to \cite{Gutperle:2016gfe}. A sketch of the regularized manifold is shown in Figure \ref{fig3}.

Once the filed $\phi_k$ is constructed we proceed to the computation of the on shell action. In particular one finds
\begin{eqnarray}
\frac{2 }{\eta \mathrm{Vol}{S^{d-1}} \ell^{d-1}}\delta I_k&=&\int_{\epsilon}^{\infty}\left(\frac{\cosh y}{\sinh p}\right)^2 \partial_y \phi_k^2 \big|_{y=y_\infty} dp-\int_{\epsilon}^{\infty}\left(\frac{\cosh y}{\sinh p}\right)^2 \partial_y \phi_k^2 \big|_{y=-y_\infty} dp \nonumber\\
& &-\int_{-y_\infty}^{y_\infty}\partial_p \phi_k^2 \big|_{p=\epsilon} dy.
\end{eqnarray}

Putting all the contributions together gives
\bea\label{eq:bulk QIM}
{ G}_{\lambda\lambda}^\epsilon&=& \frac{\mathrm{Vol}{S^{d-1}}}{2}{\eta \ell^{d-1}} \bigg[ \int^\infty_\epsilon dp \frac{\cosh^2 y}{\sinh^2 p} \( \partial_y \phi_2^2- \partial_y \phi_1^2
\)|_{y= y_\infty}
\nonumber \\
& &~-\int^{y_\infty}_{-y_\infty} dy \( \partial_p \phi_2^2- \frac{1}{2}\partial_p \phi_1^2
\)\bigg|_{p= \epsilon} \bigg]
\eea
At this point we can safely take the $y_\infty \rightarrow \infty$ limit.
The result will be divergent as $\epsilon \rightarrow 0$, however it is going to contain a universal term. The universal term will be the finite term for $d$ even and the coefficient of the logarithmic divergence for $d$ odd. In all cases we will in general be able to subtract the power divergences by the use of counterterms. Since the cut off surface is more complicated than the one usually used for holographic normalization a rigorous derivation of the construction of the counterterms is not available. However notice that our set up is similar to an interface field theory. Since the counterterms are used to regulate the divergences associated with the interface one would expect that the counterterms are localized on the $p=\epsilon$ surface. This surface preserves some of the bulk symmetries, it is then natural to look for counterterms that respect this symmetry. If one tries to construct such counterterms one would discover that they involve only odd powers of $\epsilon$. For this reason it is safe to assume that a minimal regularization scheme can be performed also in the bulk. 
\subsection{Explicit examples}
In this section we explicitly construct a couple of examples to show how the 
quantum 
information metric can be obtained holographically.

\subsubsection{$d=2$ $\Delta=6$}
We can evaluate the integral appearing in equation (\ref{eq:integration to get the field}). The result is more conveniently expressed in the coordinates of equation (\ref{eq:AdS slicing}). We find
\begin{eqnarray}
\phi_1&=& \cosh ^4 y  \, \text{csch}^4 p +\cosh ^2 y \, \text{csch}^2 p+\frac{1}{6} \nonumber \\
\phi_2&=&\frac{e^{y } \, \text{sech}^5 y \, \text{csch}^4 p}{1536} \bigg(3 \left(e^{2 y }+e^{4 y }+6\right) \left(6 e^{2 y }+e^{4 y }+11\right)+\nonumber\\
& &+2 \left(53 e^{2 y }+19 e^{4 y }+3 e^{6 y }+85\right) \cosh 2 p +\left(5 e^{2 y }+e^{4 y }+10\right) \cosh  4 p \bigg)
\end{eqnarray}
We can now proceed and plug the expression for the filed in equation (\ref{eq:bulk QIM}). In the $\epsilon\rightarrow 0$ limit we find that there is a constant cut off independent term:
\begin{equation}
G_{\lambda \lambda}=-\frac{\pi  \eta \ell}{180} 
\end{equation}
which agrees with the result derived in equation (\ref{eq:QIM d=2 examples}).
In the same way it is quite easy to find the quantum 
information metric holographically when both $d$ and $\Delta$ are even \footnote{The main obstruction when either $d$ or $\Delta$ is odd is to solve the integral of equation (\ref{eq:integration to get the field}).}. We performed this computation for various cases finding always perfect agreement with equation (\ref{eq:G-d-even}).

\subsection{Marginal deformation}
In case of a marginal deformation ($\Delta=d$) we can find the expression of the quantum information metric for any dimension $d$. The reason for it is that since the coupling is marginal the source does not transform when changing coordinates. This makes the integral easier. One finds the following expression for the quantum information metric\footnote{For a detailed derivation we refer to \cite{Trivella:2016brw}.}
\begin{equation}
G_{\lambda \lambda}^\epsilon=\frac{\eta \, \Gamma\left(\frac{1+d}{2}\right) \ell^{d-1}}{  
\Gamma(d/2)} \frac{2 \pi^{d/2}}{\Gamma(d/2)} \int_{0}^{1/\epsilon}\frac{r^{d-1}}{\sqrt{1+r^2}}dr.
\end{equation}
We want to extract the universal contribution of this quantity.
\subsubsection{d odd}
For $d$ odd this term has a logarithmic divergence. So we can look at minus the coefficient of the $1/\epsilon$ divergence of $G_{\lambda \lambda}$.
\begin{eqnarray}\label{eq: godd scalar}
G_{\lambda \lambda}^{\log}&=&\frac{\eta \, \Gamma\left(\frac{1+d}{2}\right)\ell^{d-1}}{ 
\Gamma(d/2)} \frac{2 \pi^{d/2}}{\Gamma(d/2)} \text{Res}\left(\frac{\epsilon ^{-d}}{\sqrt{\epsilon ^2+1}}\right)\bigg|_{\epsilon=0} \nonumber\\
&=& 2 \eta \ell^{d-1} \frac{(-1)^{(d-1)/2} \pi^{d/2-2}}{  \Gamma(d/2)} 
\end{eqnarray} 
where we used
\begin{equation}
\text{Res}\left(\frac{\epsilon ^{-d}}{\sqrt{\epsilon ^2+1}}\right)\bigg|_{\epsilon=0}=\frac{(-1)^{(d-1)/2} \Gamma\left(\frac{d}{2}\right)}{\sqrt{\pi}\Gamma\left(\frac{d+1}{2}\right)}.
\end{equation}
Equation (\ref{eq: godd scalar}) matches equation (\ref{eq:QIM marginal}).
\subsubsection{d even}
For $d$ even we have
\begin{equation}
\int_{0}^{1/\epsilon}\frac{r^{d-1}}{\sqrt{1+r^2}}dr=\frac{1}{2} (-1)^{-d/2} B_{-\frac{1}{\epsilon ^2}}\left(\frac{d}{2},\frac{1}{2}\right),
\end{equation}
where $B_{-\frac{1}{\epsilon ^2}}\left(\frac{d}{2},\frac{1}{2}\right)$ indicate the incomplete beta function.
We now need to isolate its constant term. To do that we express the incomplete beta function in terms of the hypergeometric function
\begin{equation}
B_{z}(A,B)=\frac{z^A \, _2F_1(A,1-B;A+1;z)}{A}
\end{equation}
and we use the following property of the hypergeometric function:
\begin{eqnarray}
\, _2F_1(a,b;c;z)&=&\frac{(-z)^{-a} \Gamma (c) \Gamma (b-a) \, _2F_1\left(a,a-c+1;a-b+1;\frac{1}{z}\right)}{\Gamma (b) \Gamma (c-a)}\nonumber \\
& &+\frac{(-z)^{-b} \Gamma (c) \Gamma (a-b) \, _2F_1\left(b,b-c+1;-a+b+1;\frac{1}{z}\right)}{\Gamma (a) \Gamma (c-b)}.
\end{eqnarray}
Using these equations with $z=-\epsilon^{-2},A=n, B=1/2, a=A, b=1-B, c=A+1$ gives the following result
\begin{equation}
\int_{0}^{1/\epsilon}\frac{r^{d-1}}{\sqrt{1+r^2}}dr=\frac{\Gamma \left(\frac{1}{2}-\frac{d}{2}\right) \Gamma \left(\frac{d}{2}\right)}{2 \sqrt{\pi }}+\frac{\epsilon ^{1-d} \, _2F_1\left(\frac{1}{2},\frac{1}{2}-\frac{d}{2};\frac{3}{2}-\frac{d}{2};-\epsilon ^2\right)}{d-1}.
\end{equation}
We can now expand the hypergeometric function for small $\epsilon$. Since $d$ is even the second term will correspond to a Laurent expansion with only odd powers of $\epsilon$. Thus the only finite part is the first contribution. We are then left with:
\begin{equation}
G_{\lambda \lambda}=\eta \ell^{d-1} \frac{(-1)^{\frac{d}{2}} \pi ^{\frac{d}{2}}}{ \Gamma \left(\frac{d}{2}\right)}
\end{equation}
which agrees with equation (\ref{eq:QIM marginal}).
\section{Information metric and interface free energy of conformal Janus on $S^{d}$}\label{sec: free energy}
In this section we relate the quantum information metric on the cylinder to the free energy of a conformal Janus configuration on the Euclidean sphere.

As usual we start with the expression of the overlap between the deformed ground state and the undeformed one:
\begin{equation}
\braket{\Psi_1|\Psi_0}=\frac{Z_2}{\sqrt{Z_1 Z_0}}
\end{equation}
We can compute the $Z_k$ holographically by $Z_k=\exp(-I_k)$ where $I_k$ is the on shell action of a Einstein-dilaton theory. If the deformation is marginal we have
\begin{equation}
Z_0=Z_1=\exp(-I_{AdS}),
\end{equation}
and thus
\begin{equation}
\braket{\Psi_1|\Psi_0}=\exp(-(I_2-I_{AdS})).
\end{equation}
If we expand the left hand side for small $\delta \lambda$ we have
\begin{equation}
\braket{\Psi_1|\Psi_0}=1-G_{\lambda \lambda}\delta \lambda^2+\mathcal{O}(\delta \lambda^3) 
\end{equation}
Thus
\begin{equation}
\log(\braket{\Psi_1|\Psi_0})=-G_{\lambda \lambda}\delta \lambda^2+\mathcal{O}(\delta \lambda^3)
=-(I_2-I_{AdS}),
\end{equation}
which results in
\begin{equation}
\Delta F=G_{\lambda \lambda}\delta \lambda^2+\mathcal{O}(\delta \lambda^3).
\end{equation}
Therefore the free energy of a Janus interface at second order in the Janus deformation parameter reproduces the quantum information metric for a CFT ground state living on $\mathbb{R}\times S^{d-1}$.

At this point we want to relate the computation of the quantum information metric on $\mathbb{R}\times S^{d-1}$  to the computation of the free energy on $S^d$.
We can map the cylinder to the sphere. A way to do this is to take the cylinder with metric
\begin{equation}
ds^2_{cyl}=d\tau^2+ds^2_{S^{d-1}}
\end{equation}
and conformally map it to a sphere with metric
\begin{equation}
ds_{S^d}^2=d\theta^2+\sin^2\theta  ds^2_{S^{d-1}}
\end{equation}
by using the following change of coordinates
\begin{equation}\label{eq:confmap}
\tau=\log(\tan(\theta/2)).
\end{equation}
We are allowed to perform the change of coordinates because the fields well behave at $\tau=\pm \infty$. We will return on this detail later. Under the map (\ref{eq:confmap}) the interface at $\tau=0$ is mapped to the equator of the sphere, the $\tau>0$ ($<0$) region is mapped to the northern (southern) hemisphere and the cut off surfaces $\tau=\pm \epsilon$ are mapped to cut off surfaces located at constant $\theta=2 \arctan(e^{\pm \epsilon})$.

To find the quantum information metric on the cylinder one has to compute 
\begin{equation}
\int_{\tau_1>\epsilon}\int_{\tau_2<-\epsilon}\braket{\mathcal{O}(\tau_1,\Omega_1)\mathcal{O}(\tau_2,\Omega_2)}.
\end{equation}
Under the conformal transformation (\ref{eq:confmap}) this maps to
\begin{equation}
\int_{\tilde N}\int_{\tilde S}\braket{\mathcal{O}(\theta_1,\Omega_1)\mathcal{O}(\theta_2,\Omega_2)},
\end{equation}
where $\tilde N$  ($\tilde S$) indicates the (regularized) northern  (southern) hemisphere
Using a path integral construction we could have derived this formula by looking at the second order contribution in $\delta \lambda$ of $\Delta F_{\text{sphere}}$. This indeed shows that we can compute the quantum information metric for a marginal deformation by looking at the leading order contribution of the interface free energy.

This result can be checked analytically in the bulk. The interface free energy for the conformal Janus on the Euclidean sphere $S^d$ has been computed in \cite{Bak:2016rpn} for $d=2,3$ and indeed the small $\delta \lambda$ behavior matches the computation of the quantum information metric presented in this paper.

One could wonder if the same procedure can be applied for the quantum information metric of a CFT living on $\mathbb{R}\times \mathbb{R}^{d-1}$. In this case the interface is a codimension one plane. A conformal transformation between this configuration and a sphere with interface extended along the equator is available. Before performing the conformal map one has to compactify the space. This is not possible in this set up. The reason is that the interface extends to infinity, thus the fields generally speaking would have a non trivial behavior at large distances. We cannot therefore make the manifold compact. 

The argument explained in this section fails if the deformation is not marginal. For a non marginal deformation the conformal transformation will change the effective source. Therefore the usual configuration on the cylinder would be mapped to a configuration on the sphere with a coupling constant that depends on the polar angle.

We conclude the section with a comment about regularization.  
On the cylinder the regularization is performed by excluding from the path integral the region close to the interface: we put cut offs at $\tau=\pm \epsilon$. These cut off surfaces are mapped to $\theta=2 \arctan(e^{\pm \epsilon})\approx \pi/2\pm\epsilon$, which looks appealing since it is the natural cut off one would use. However it is important to stress that generically one should make use of the entire expression $\theta=2 \arctan(e^{\pm \epsilon})$ since the relation between the cut offs in the two geometries is non linear and thus truncating the relation for small $\epsilon$  could suppress some potential finite contributions.


\section{Concluding remarks} \label{sec:conclusions}
In this note we compute the quantum information metric for the ground state of a CFT living on $\mathbb{R} \times {S}^{d-1}$  perturbed by a scalar primary operator.  We find that when ${2\Delta-d+1}$ is even the renormalized quantum information metric
becomes anomalous depending on the radius of the sphere, explicitly breaking the scale symmetry of the underlying CFT,
otherwise it is finite and scale independent.   For integral values of $\Delta\, (\,\, > \frac{d+1}{2})$ 
we present an explicit formula for the quantum information metric, which is verified for various cases both by gravity and field theoretic computations. The renormalized quantum information metric is well defined  physically and can be measured experimentally in 
principle.  Since we now have definite predictions for the quantum information metric, our results can be used to clarify 
 a possible relation between the quantum information metric and  other quantities 
like quantum complexity  \cite{Brown:2015bva}-\cite{Alishahiha:2015rta}. 

We find that the degrees of freedom responsible for the quantum information metric are organized in a $d-1$ dimensional theory that may be viewed as localized in
the interface  ${S}^{d-1}$. A similar observation holds for the conformal Janus  on Euclidean ${S}^d$ whose interface is 
given by the equatorial sphere  ${S}^{d-1}$ \cite{Bak:2016rpn}. In that case the interface contribution of the free energy shows the characteristics of $d-1$ field theory living on the interface  ${S}^{d-1}$. 
We showed a precise match between these two different observations, in particular we established that the quantum information metric for an exactly marginal deformation reproduces the leading term of the interface free energy  of 
the conformal Janus. It will be interesting to see if this match can be generalized to the cases of non marginal operators near  the critical point.
Further investigation is  required in this direction. 

\section*{Acknowledgements}
The work of D.B. is
supported in part by
NRF Grant 2017R1A2B4003095. The work of A.T. is supported in part by the National Science Foundation under grant PHY-16-19926. We would like to thank Eric D'Hoker for careful reading of the draft.

\appendix

\section{Two point function normalization}\label{app}
In this appendix we explain our normalization of the two point function of operators that is consistent with our gravity description.
\subsection{Normalization of Boundary-Bulk propagator}
The bulk to boundary propagator in Poincar\'e coordinates is given by
\begin{equation}
K(x;x',z)=c_{\Delta} \frac{z^\Delta}{(z^2+|x-x'|^2)^{\Delta}}.
\end{equation}
The constant $c_{\Delta}$ is fixed by requiring that as $z \rightarrow 0$ one has
\begin{equation}
K(x;x',x)=z^{d-\Delta}\delta^d(x-x').
\end{equation}
We then have
\begin{equation}
1=\int d^dx z^{\Delta-d} K(x;x',z)=c_{\Delta}\int d^dx \frac{z^{2 \Delta-d}}{(z^2+|x-x'|^2)^{\Delta}}.
\end{equation}
At this point we change variable of integration by defining $x-x'=z y$, obtaining
\begin{equation}
1=c_{\Delta} \int d^dy (1+y^2)^{-\Delta}
\end{equation}
from which we find
\begin{equation}
c_{\Delta}=\left(\mathrm{Vol}{S^{d-1}}\frac{\Gamma \left(\frac{d}{2}\right) \Gamma \left(\Delta -\frac{d}{2}\right)}{2 \Gamma (\Delta )}\right)^{-1}.
\end{equation}
This leads to (\ref{cdd}).
\subsection{Normalization constant of the two point function}
The two point function on the CFT side is given by
\begin{equation}
\braket{\mathcal{O}(x_1)\mathcal{O}(x_2)}=\frac{{\cal N}_\Delta}{|x_1-x_2|^{2\Delta}}.
\end{equation}
The constant ${\cal N}_\Delta$ has to be chosen such that $\braket{\exp({\int J \mathcal{O}})}_{\mathrm{CFT}}=\exp{(-I_{\mathrm{bulk}})}$, where $I_{\mathrm{bulk}}$ is the on shell action of the bulk theory. We consider for the gravity side a free massive scalar. We have that, given the boundary condition $J$, the field is reconstructed in the bulk using the boundary to bulk propagator:
\begin{equation}
\phi(x,z)=c_\Delta \int d^dx' \frac{z^\Delta J(x')}{(z^2+|x-x'|^2)^{\Delta}}.
\end{equation}
As we approach the boundary we have that the leading contribution is given by 
\begin{equation}
K(x;x',z)=\begin{cases}
z^{d-\Delta} \delta^d(x-x') &\text{if  } x=x'\\
c_{\Delta} \frac{z^\Delta}{|x-x'|^{2\Delta}}. &\text{if  } x\neq x'.
\end{cases}
\end{equation}

This means that we have the following expansion for the field close to the boundary
\begin{equation}
\phi= J(x) z^{d-\Delta}+\cdots +c_\Delta z^\Delta \int d^dx'\frac{J(x')}{|x-x'|^{2 \Delta}}+\cdots 
\end{equation}
The first part of this equation involves only local terms, as we know that for $x\neq x'$ $K \approx z^{\Delta}$. In the following we will need also $\partial z \phi$, we have
\begin{equation}
\partial_z\phi=(d-\Delta) J(x) z^{d-\Delta-1}+\cdots +c_\Delta \Delta z^{\Delta-1} \int d^dx'\frac{J(x')}{|x-x'|^{2 \Delta}}+\cdots
\end{equation}
The Euclidean action is given by
\begin{equation}
I_{\mathrm{bulk}}=\eta \int d^{d}x dz \sqrt{g} (\partial_\mu \phi \partial_\nu \phi g^{\mu \nu}+m^2 \phi^2) 
\end{equation}
We regularize it by putting a cut off at $z=\epsilon$. Integrating by parts and using the equations of motion gives
\begin{equation}
I_{\mathrm{bulk}}=-\eta \int_{z=\epsilon} d^{d}x (\sqrt{g}g^{zz}\phi \partial_z \phi ).
\end{equation}
The finite part of the on shell action is 
\begin{eqnarray}
I_{\mathrm{bulk}}&=&-\eta \int_{z=\epsilon} d^{d}x\left( J(x) c_\Delta \Delta \int d^dx' \frac{J(x')}{|x-x'|^{2\Delta}}+J(x) c_\Delta (d-\Delta) \int d^dx' \frac{J(x')}{|x-x'|^{2\Delta}}\right)\nonumber \\
&=&\eta c_\Delta d  \int_{z=\epsilon} d^{d}x d^dx' \frac{J(x')J(x)}{|x-x'|^{2\Delta}}.
\end{eqnarray}
Assuming that the counter terms do not change the finite part of the action (which we know to be true for our set up, even if one has a non trivial cut off surface) we have
\begin{equation}\label{C new}
{\cal N}_\Delta=2 \eta\ell^{d-1} c_\Delta d =  \frac{2 \eta\ell^{d-1}  d \, \Gamma (\Delta )}{ \pi^{d/2} \Gamma \left(\Delta -\frac{d}{2}\right)}
\end{equation} 
The normalization of two-point function  was computed in  \cite{Freedman:1998tz}, whose  result disagrees with ours  by an extra factor of 
$\Delta/d$. Note that the authors of \cite{Freedman:1998tz} suggested that the normalization they used needed a modification for $\Delta \neq d$.

\end{document}